\documentclass[prl,aps,twocolumn]{revtex4}
\usepackage{epsfig}
\usepackage{bm}

\begin{document}


\title{Josephson Current in S-FIF-S Junctions: \\
Nonmonotonic Dependence on Misorientation Angle}
\author{Yu.~S.~Barash$^{1}$, I.~V.~Bobkova$^{1}$ and T. Kopp$^{2}$}
\affiliation{$^1$Lebedev Physical Institute, Leninsky Prospect
53,\, Moscow 119991, Russia\\ $^2$Center for Electronic
Correlations and Magnetism,
Institute of Physics, University of Augsburg, D-86135 Augsburg, Germany\\
}




\begin{abstract}
Spectra and spin structures of Andreev interface states in {\it
S-FIF-S} junctions are investigated with emphasis on finite transparency and
misorientation angle $\varphi$ between in-plane magnetizations of
ferromagnetic layers in a three-layer interface. It is demonstrated
that the Josephson current in {\it S-FIF-S} quantum point contacts can
exhibit a nonmonotonic dependence on the misorientation angle.
The characteristic behavior takes place, if the $\pi$-state is the equilibrium
state of the junction in the particular case of parallel magnetizations.
\end{abstract}


\maketitle





The dc Josephson effect in junctions with ferromagnetic interfaces
exposes remarkable features which have been intensively studied in
recent years theoretically
\cite{bul77,buz82,mrs88,buz92,demler97,yip00,heik00,fogel00,%
fogel01,koshina011,koshina012,volkov01,nazarov01,bb02,waintal02,gol021,gol022}
and experimentally \cite{ryazanov01,kontos02}. Apart from interfaces
with a fixed magnetization, considerable attention has been drawn also
to more complicated cases, when the magnetization is spatially
dependent inside the interface. An important particular model for this
kind of interfaces is a three-layer {\it FIF}-interface, where two metallic
ferromagnetic layers with in-plane magnetizations, making angle
$\varphi$ with each other, are separated by an insulating magnetically
inactive interlayer
\cite{koshina011,koshina012,volkov01,waintal02,gol021}. In the present
article we identify theoretically spectra and spin polarization of
Andreev states bound to the three-layer {\it FIF}-constrictions with finite
transmission, separating clean $s$-wave superconductors. Then we
determine the Josephson current in the {\it S-FIF-S} quantum point
contacts. This problem was not studied previously in the literature.  In
the dirty limit, considered in \cite{koshina011,koshina012,gol021},
Andreev bound states are fully destroyed. Considerations of
Ref.\,~\onlinecite{volkov01} imply the absence of Andreev bound states
in clean {\it S-FIF-S} junctions. This can be justified only for short
superconductors, whose lengths are less than their coherence lengths.
Spectra of Andreev states and the Josephson current in {\it S-FIF-S}
junctions with $\varphi\ne0$ have been found earlier only in the
particular limit of fully transparent constriction which was
characterized by the authors as a toy model \cite{waintal02}.

The misorientation angle $\varphi$ can be considered, in general, as a
variable quantity. Let, for instance, the magnetization axis be pinned
in one ferromagnetic layer, while in the other one there is an easy
in-plane magnetization layer. Then one can vary the misorientation
angle (keeping other parameters fixed) by
applying an external weak magnetic field to the interface.  We find
that the Josephson current as a function of the misorientation angle
$\varphi$ manifests a characteristic nonmonotonic behavior, if, at
$\varphi=0$, the $\pi$-state is the equilibrium state of the junction.

For our analysis, we examine a smooth plane interface between two
superconductors which consists of two layers of the same ferromagnetic
metal separated by an insulating nonmagnetic barrier. Two identical
ferromagnetic layers are characterized by their thickness $l$ and
internal exchange fields $|\bm{h}_{1,2}|=h$, which, being parallel to
the layers, make an angle $\varphi$ with each other.

{\it The normal-state scattering $\cal S$ matrix} is represented as
${\cal S}=S(1+\hat\tau_z)/2+\widetilde S(1-\hat\tau_z)/2$, where the
Pauli-matrices $\hat\tau_j$ are taken in particle-hole space and
$\widetilde S(p_\|)=S^{T}(-p_\|)$.  Each component $\hat{S}_{ij}$ in
matrix $S=\|\hat{S}_{ij}\|$ ($i(j)=1,2$) is in its turn a matrix in
spin space.  Matrix $\hat{S}_{ii}=\left(\begin{array}{cc}
r_{i}^{\uparrow\uparrow} &r_{i}^{\uparrow\downarrow} \\
 r_{i}^{\downarrow\uparrow}&r_{i}^{\downarrow\downarrow}
\end{array}\right)$ contains, in general, spin-dependent interface
reflection amplitudes for normal-state quasiparticles in $i$-th
half-space, while $\hat{S}_{ij}=\left(\begin{array}{cc}
d_{ij}^{\uparrow\uparrow} &d_{ij}^{\uparrow\downarrow} \\
 d_{ij}^{\downarrow\uparrow}&d_{ij}^{\downarrow\downarrow}
\end{array}\right)$ with $i\ne j$ incorporates spin-dependent
transmission amplitudes for normal-state quasiparticles from side $j$.
For the interface potentials conserving particle current, the
scattering matrix has to satisfy the unitarity condition: ${\cal
S}{\cal S}^\dagger=1$. If the interface Hamiltonian possesses
time-reversal symmetry, one obtains an additional constraint on the
scattering matrix: $ S(\bm{p}_f, \bm{h}_{1,2})=\hat \sigma_y
S^{T}(-\underline{\bm{p}}_f,-\bm{h}_{1,2})\hat \sigma_y$ \cite{mrs88}.
Assume, for simplicity, the exchange fields to be much smaller
compared to the Fermi energies. For the diagonalization
of the $S_{11}$-matrix we choose the $z$-axis along the magnetization
in the first (left) ferromagnetic layer. Then the other
$S_{ij}$-matrices are nondiagonal unless $\varphi=0, \pi$:
\begin{eqnarray}
\nonumber
&&\hat{S}_{21}=d\exp\left(\frac{i\Theta}{4}
\displaystyle\left(\hat{\displaystyle
\sigma}_y\sin\varphi+\hat{\displaystyle\sigma}_z\cos\varphi\right)\right)
\exp\left(\frac{i\Theta}{4}\hat{\displaystyle\sigma}_z\right)
,
\\
\nonumber
&&\hat{S}_{12}=d\exp\left(\frac{i\Theta}{4}\hat{\displaystyle\sigma}_z\right)
\exp\left(\frac{i\Theta}{4}\displaystyle\left(\hat{\displaystyle
\sigma}_y\sin\varphi+\hat{\displaystyle\sigma}_z\cos\varphi\right)\right)
,
\\
\nonumber
&&\hat{S}_{11}=r\exp\bigl(i\frac{\Theta}{2}\hat{\displaystyle\sigma}_z\bigr)
,
\\
&&\hat{S}_{22}=\tilde{r}\exp\biggl(i\frac{\Theta}{2}\displaystyle\left(\hat{
\displaystyle\sigma}_y\sin\varphi+\hat{\displaystyle\sigma}_z\cos\varphi\right)
\biggr)\enspace .
\label{s}
\end{eqnarray}
Here $\Theta=\frac{\displaystyle 4lh}{\displaystyle
\hbar {\rm v}_{f,x}}$. Quantities $r$, $\tilde r$ and $d$ are the
respective reflection and transmission amplitudes of the potential
barrier $V$, satisfying the condition $rd^*=-d\tilde{r}^*$.

We carry out calculations within the quasiclassical theory of
superconductivity, based on the equations for the so-called Riccati
amplitudes or coherence functions \cite{schopohl98,eschrig00,fogel00}.
Considering a quantum point contact with {\it FIF}-constriction, the order
parameter is taken spatially constant. We include interface exchange
fields in the quasiclassical boundary conditions. Since they imply, as
usually, that all interface potentials are much larger than the
superconducting order parameter \cite{zai84}, one should assume
$|h_{1,2}|\gg{\mit \Delta}$.

With the above normal-state $\cal S$ matrix we get {\it four branches of
interface Andreev bound states}:
\begin{equation}
\varepsilon_{1,2}=|{\mit\Delta}|\cos\frac{\Phi_{1,2}}{2} \enspace ,
\qquad \varepsilon_{3,4}=-|{\mit\Delta}|\cos\frac{\Phi_{1,2}}{2}
\enspace , \label{cd}
\end{equation}
where the quantities $\Phi_{1,2}(\chi,\Theta,\varphi)$ are
defined as
\begin{eqnarray}
\nonumber
\cos\Phi_{1,2}(\chi,\Theta,\varphi)=\cos\Theta -2D\cos\Theta\sin^2\frac{
\displaystyle\chi}{\displaystyle2}+ \qquad\quad\\
\nonumber
+2D\cos\chi\sin^2\frac{\displaystyle\Theta}{
\displaystyle2}\sin^2\frac{\displaystyle\varphi}{\displaystyle2}
\pm 2\sqrt{D}\sin\frac{\displaystyle\chi}{\displaystyle2}
\sin\Theta\cos\frac{\displaystyle\varphi}{\displaystyle2}\times \\
\times\sqrt{1-D\sin^2\frac{
\displaystyle\chi}{\displaystyle2}+D\cos^2\frac{\displaystyle\chi}{
\displaystyle2}\tan^2\frac{\displaystyle\Theta}{\displaystyle2}
\sin^2\frac{\displaystyle\varphi}{\displaystyle2}}
\enspace .
\label{phi}
\end{eqnarray}
Here, $\chi$ is the phase difference of the two superconductors.
The energies $\varepsilon_{i}$ ($i=1,2,3,4$)
implicitly depend on quasiparticle momentum directions via the
parameter $\Theta$ and the transmission coefficient $D$.

For $\varphi=0$ the spectra Eq.~(\ref{phi}) reduce to those found in
Ref.\,~\onlinecite{fogel01}. In the particular case of antiparallel
orientation of the left and the right magnetization $\varphi=\pi$, the
spectra of Andreev interface states (\ref{cd}), (\ref{phi}) take the
form
\begin{equation}
\varepsilon_{1}=\varepsilon_{2}=-\varepsilon_{3}=-\varepsilon_{4}=
|{\mit\Delta}|\sqrt{D\cos^2\frac{\displaystyle\chi}{\displaystyle2}+R\cos^2
\frac{\displaystyle\Theta}{\displaystyle2}}
\enspace .
\label{pi}
\end{equation}
Being symmetric with respect to the transformation $\Theta\to- \Theta$,
the spectrum (\ref{pi}) is doubly degenerated.  In the limit of a
nonmagnetic interface ($\Theta=0$), our result, Eqs.~(\ref{cd}) and
(\ref{phi}), leads to a well known spectrum of spin-degenerate
interface Andreev bound states \cite{fur90,fur191,been191,been291}
$\varepsilon_B=\pm|{\mit\Delta}|\sqrt{1-D\sin^2(\chi/2)}$.

Quasiparticle spin is a qood quantum number in the BCS theory of
superconductivity, if one can disregard spin-flip effects stimulated,
for instance, by magnetic impurities, spin-orbit interactions or
magnetically active interfaces. In the presence of a paramagnetic spin
interaction with an externally applied magnetic field or an internal
exchange field, spin degeneracy of quasiparticle states is lifted and
only states with parallel or antiparallel spin-to-field orientations
are still eigenstates of the problem. This can lead to effects having
physics common with the Larkin-Ovchinnikov-Fulde-Ferrell state
\cite{ff65,lo65} and, in particular, associated with opposite signs of
the Zeeman terms for electrons forming a Cooper pair in singlet
superconductors.

A Bogoliubov quasiparticle in the superconductor has well defined spin,
although its electron and hole components are described with Zeeman
terms of opposite signs. Also, an electron and its Andreev reflected
partner (hole) at an interface, separating singlet superconductors and
leading to no spin-flip processes, have identical spin orientations and
opposite signs of Zeeman terms. With opposite velocity directions and
electric charges, while in identical spin states, they carry jointly
the electric supercurrent across the interface, but no equilibrium spin
current. Hence, definite spin polarization of interface Andreev bound
states is fully compatible with the fact that Cooper pairs in singlet
supercoductors carry no spin current.

Andreev states bound to nonmagnetic interfaces are spin degenerated.
For a ferromagnetic interface with uniformly oriented magnetization
Andreev interface states are spin polarized, being parrallel or
antiparallel to the magnetization. The ferromagnetic interface lifts
spin degeneracy of the Andreev states, but still does not mix the
spin-polarized channels, carrying the Josephson current \cite{bb02}.
This is not the case, however, if various orientations of magnetization
are present in the interface, as this takes place in the
{\it FIF}-interface with $\varphi\ne0$. Quasiparticle Andreev interface
states with the spectra of Eqs.~(\ref{cd}), (\ref{phi}) are characterized by
a nontrivial spin structure, which substantially depends (together with
the spectra themselves) on $\varphi$, $\Theta$ and $D$. In general,
each of the two incoming and two outgoing parts of quasiparticle
trajectories, forming an Andreev interface state, has its own specific
spin polarization. This should be compatible with no spin current
they carry, on the whole, across the interface.
\begin{figure}[!tbh]
  \begin{minipage}[b]{.5\linewidth}
   \centering \epsfig{figure=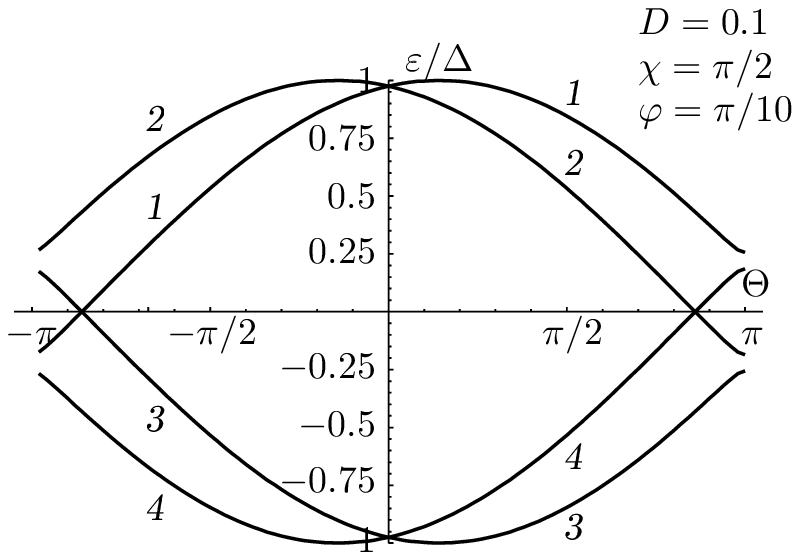,width=\linewidth}
  \end{minipage}\hfill
  \begin{minipage}[b]{.5\linewidth}
   \centering \epsfig{figure=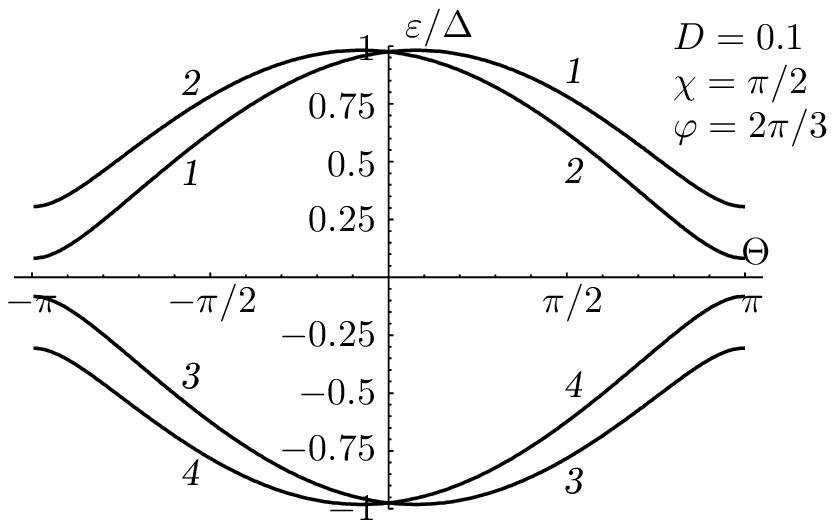,width=\linewidth}
  \end{minipage}\hfill
  \begin{minipage}[b]{.5\linewidth}
   \centering \epsfig{figure=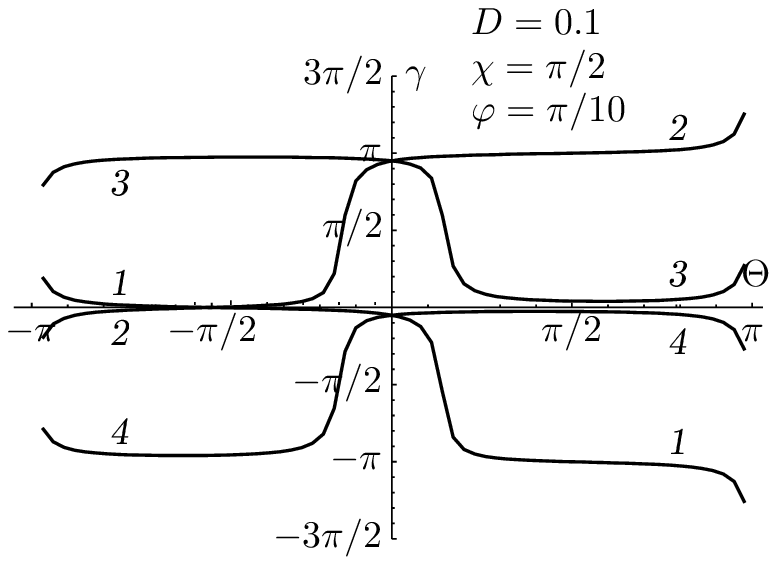,width=\linewidth}
  \end{minipage}\hfill
  \begin{minipage}[b]{.5\linewidth}
   \centering \epsfig{figure=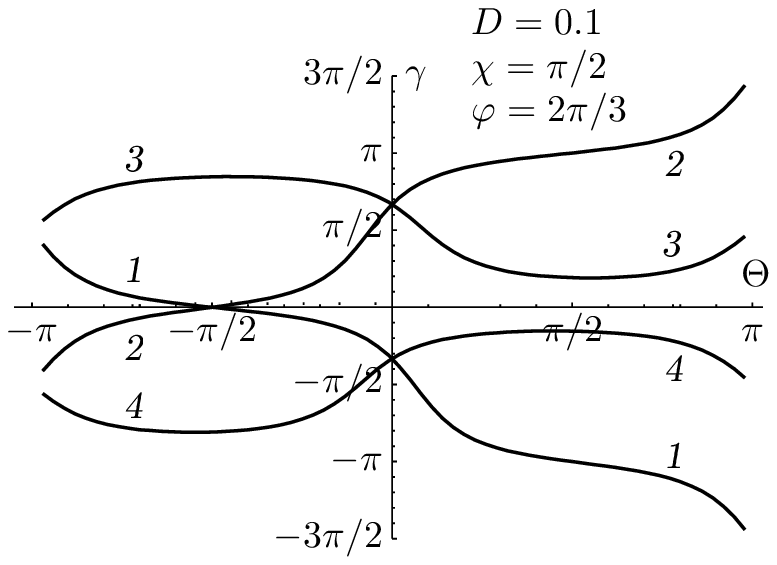,width=\linewidth}
  \end{minipage}\hfill
\caption{ Upper panel: Energies of the four branches of Andreev
interface states as functions of $\Theta$. Lower panel: Angle
$\gamma(\Theta)$ of the spin of an incoming quasiparticle in the right
superconductor with the magnetization of the right ferromagnetic layer,
for each of the four branches of the Andreev interface states. Left
column: $\varphi= 0.1\pi$. Right column: $\varphi=2\pi/3$.
Transparency and phase difference have values $D=0.1$ and $\chi=\pi/2$,
respectively. }
\label{theta01}
\end{figure}
\begin{figure}[!tbh]
  \begin{minipage}[b]{.5\linewidth}
   \centering \epsfig{figure=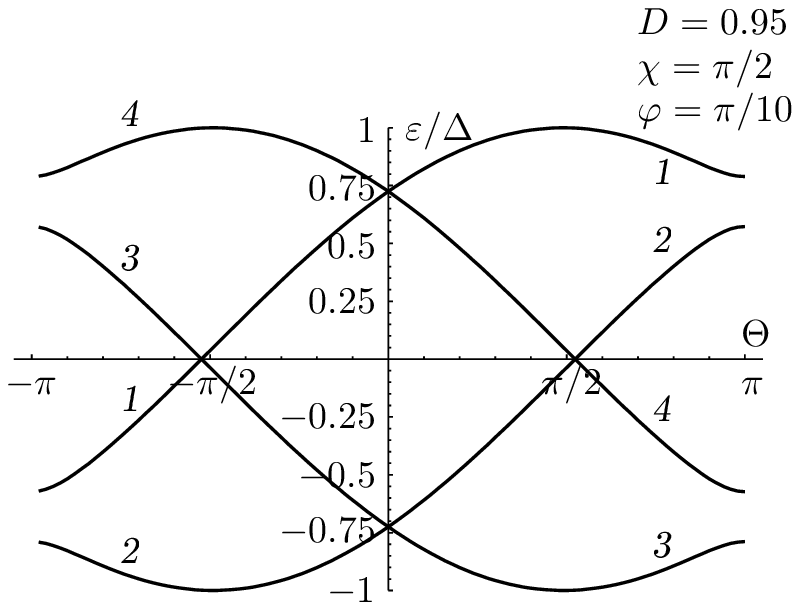,width=\linewidth}
  \end{minipage}\hfill
  \begin{minipage}[b]{.5\linewidth}
   \centering \epsfig{figure=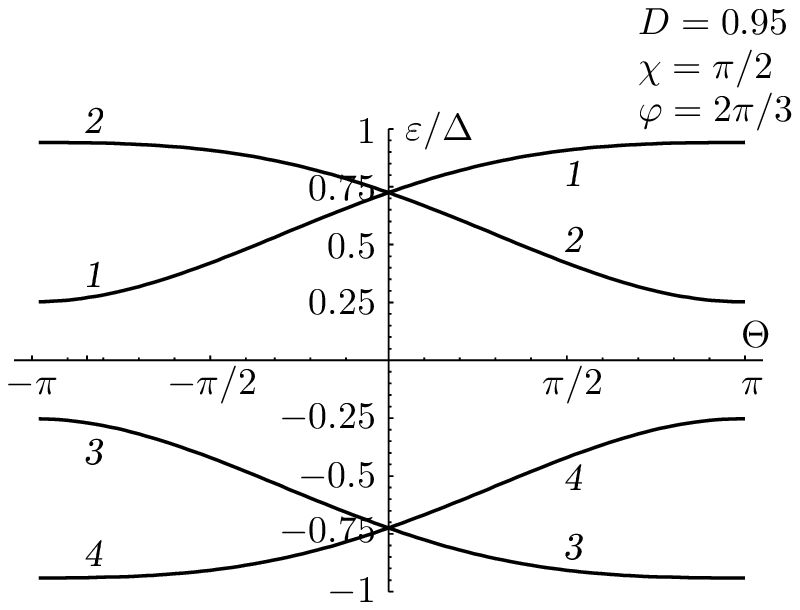,width=\linewidth}
  \end{minipage}\hfill
  \begin{minipage}[b]{.5\linewidth}
   \centering \epsfig{figure=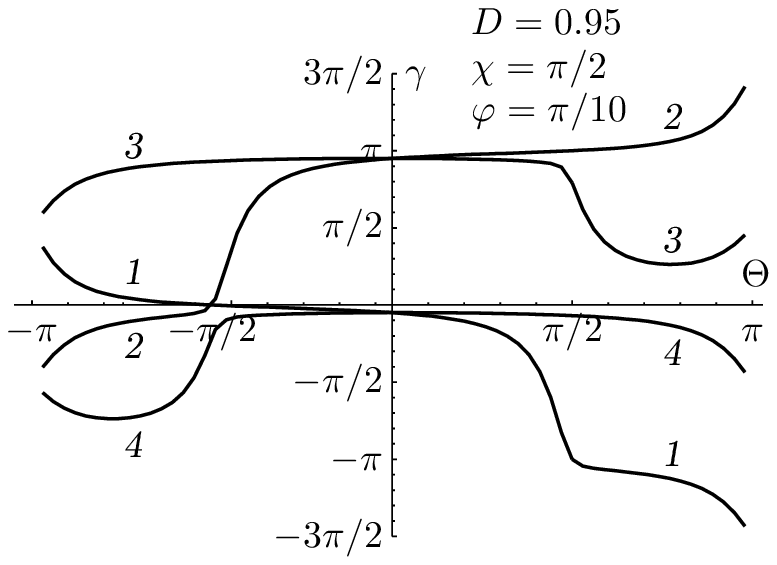,width=\linewidth}
  \end{minipage}\hfill
  \begin{minipage}[b]{.5\linewidth}
   \centering \epsfig{figure=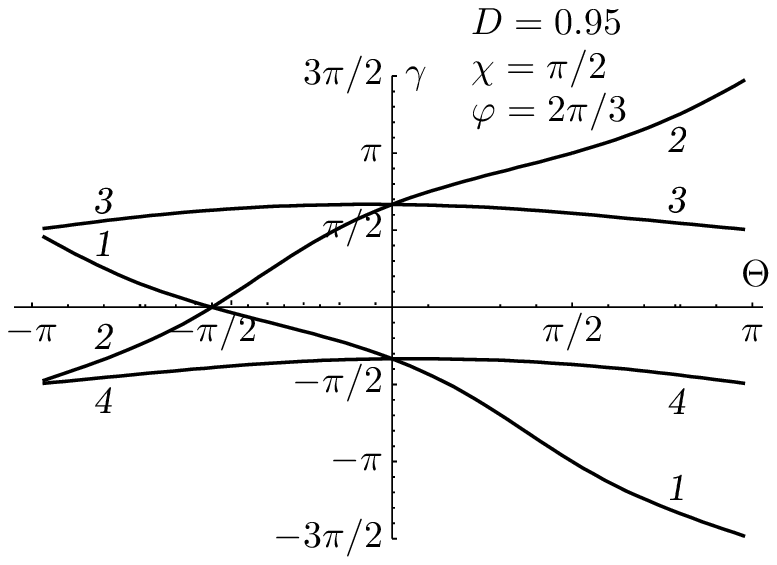,width=\linewidth}
  \end{minipage}\hfill
\caption{ The same as in Fig.\ \ref{theta01} for highly transparent
junctions with $D=0.95$.}
\label{theta095}
\end{figure}
Figs.~\ref{theta01} and \ref{theta095} demonstrate the evolution of
spectra and spin polarizations of four branches of Andreev interface
states as functions of $\Theta$ in tunnel junctions (with transparency
$D=0.1$) and highly transparent junctions ($D=0.95$) respectively. Two
particular relative orientations of magnetization $\varphi=0.1\pi$
(left column) and $\varphi=2\pi/3$ (right column) are chosen. For
definiteness, we consider spin polarizations of Andreev states on the
incoming part of the quasiparticle trajectory in the right
superconductor. The spin polarization gradually changes with the
parameter $\Theta$ in all cases considered. A characteristic scale of
$\Theta$ for the spin reconstruction diminishes with decreasing the
misorientation angle $\varphi$. For vanishing $\varphi$ the scale
vanishes and there are abrupt jumps from parallel to antiparallel (or
vice versa) spin orientations with respect to the magnetization, taking
place at those values of $\Theta$, where
$\varepsilon_i(\Theta)=\pm{\mit\Delta}$\, \cite{bb02}.

Only in the particular case $\varphi=0$, when a single magnetization direction
is present inside a symmetric magnetic interface, each of the Andreev interface
states possesses, as a whole, a definite spin-up or spin-down polarization,
identical for all incoming and outgoing quasiparticle trajectories forming the
state. Then the spectra of the spin-up
polarized Andreev bound states are \cite{bb02}: $\varepsilon^{\uparrow}_{1,2}=
|{\mit\Delta}|\,{\rm sgn}\!\left(\!\sin\frac{\displaystyle\Phi_{1,2}(\chi,\Theta,
\varphi=0)}{\displaystyle2}\!\right)\!\cos\!\frac{\displaystyle\Phi_{1,2}(\chi,
\Theta,\varphi=0)}{\displaystyle2}$. The energies
$\varepsilon^{\downarrow}_{1,2}$ for spin-down Andreev bound states
are obtained from $\varepsilon^{\uparrow}_{1,2}$ by substituting $\Theta\to-
\Theta$.

\begin{figure}[!tbh]
  \begin{minipage}[b]{.5\linewidth}
   \centering \epsfig{figure=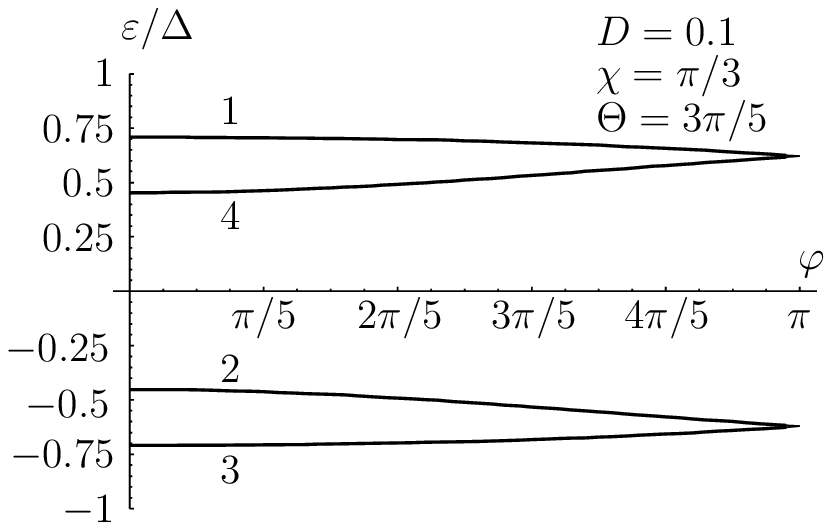,width=\linewidth}
  \end{minipage}\hfill
  \begin{minipage}[b]{.5\linewidth}
   \centering \epsfig{figure=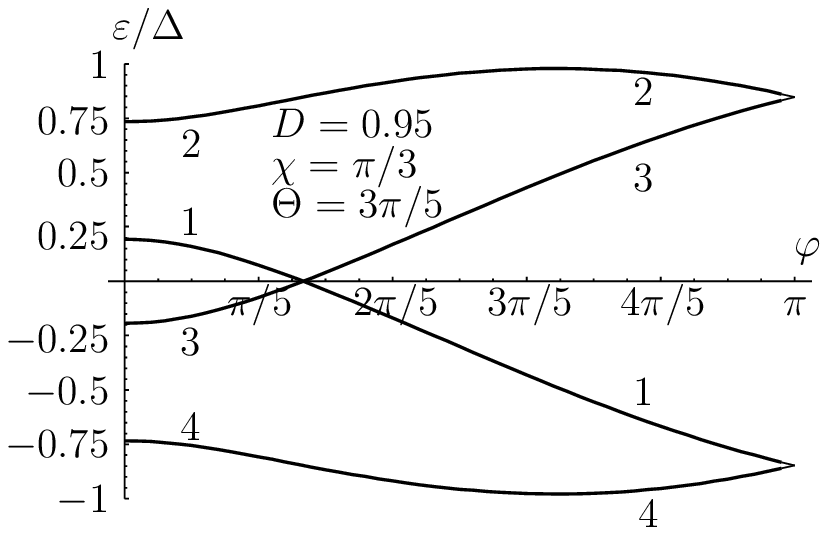,width=\linewidth}
  \end{minipage}\hfill
  \begin{minipage}[b]{.5\linewidth}
   \centering \epsfig{figure=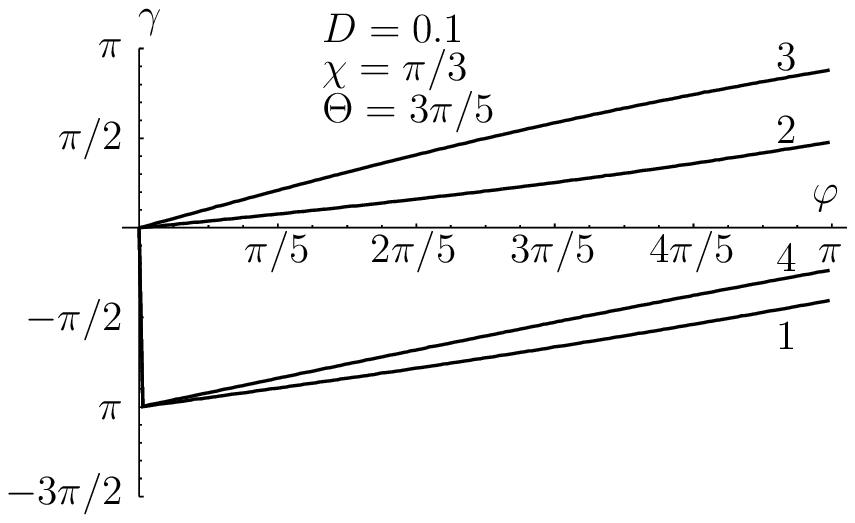,width=\linewidth}
  \end{minipage}\hfill
  \begin{minipage}[b]{.5\linewidth}
   \centering \epsfig{figure=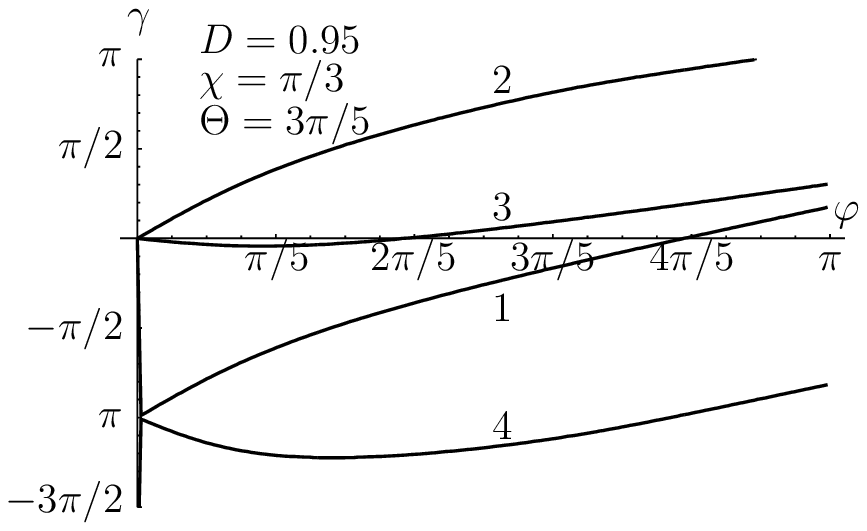,width=\linewidth}
  \end{minipage}\hfill
\caption{Upper panel: Energies of the four branches of Andreev interface states as
functions of misorientation angle $\varphi$. Lower panel: Angle
$\gamma(\varphi)$
of the spin of an incoming quasiparticle in the
right superconductor with the magnetization of the right ferromagnetic layer,
for each of four branches of the Andreev interface states.
Left (right) column: $D= 0.1$ ($D=0.95$). The phase difference is
$\chi=\pi/3$, and $\Theta=3\pi/5$.
}
\label{figphi}
\end{figure}

The spectra of Andreev states and their spin polarizations as functions
of the misorientation angle $\varphi$ are shown in Fig.~\ref{figphi}.
The spin polarization at $\varphi\ne0$ makes a finite angle with both
magnetization directions and differs on different incoming and outgoing
trajectories related by the bound state. As already mentioned above,
for antiparallel magnetizations ($\varphi=\pi$) the spectra are doubly
degenerated. Spin polarizations, shown in Fig.~\ref{figphi}
for $\varphi=\pi$, can be considered as correct eigenfunctions in
zeroth order approximation in small deviations $\varphi-\pi$.

The spin structure of Andreev interface states at nonzero
$\varphi$ should be taken into account in producing
nonequilibrium occupation of the states. For $\varphi=0$ only the
interlevel transitions accompanied with spin-flip processes are possible
under certain conditions \cite{bb02}. On account of the complicated spin
structure of the Andreev states at nonzero $\varphi$, there are
actually no strict restrictions to a change of quasiparticle spin in the
transition.

\begin{figure}[!tbh]
\begin{minipage}[b]{.5\linewidth}
   \centering \epsfig{figure=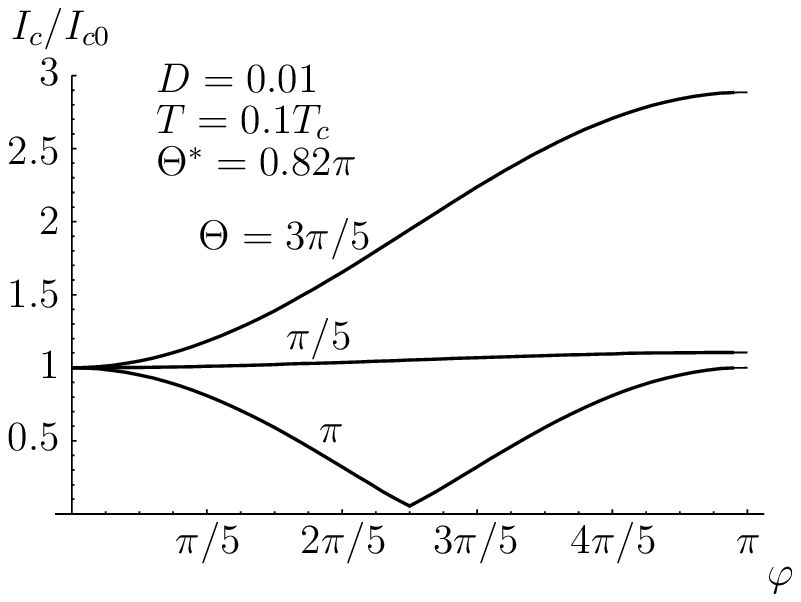,width=\linewidth}
  \end{minipage}\hfill
  \begin{minipage}[b]{.5\linewidth}
   \centering \epsfig{figure=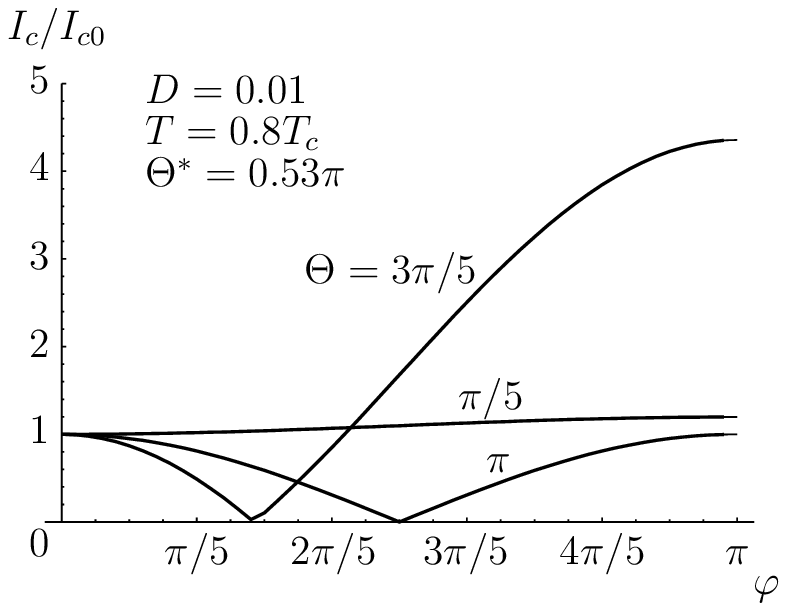,width=\linewidth}
  \end{minipage}\hfill
  \begin{minipage}[b]{.5\linewidth}
   \centering \epsfig{figure=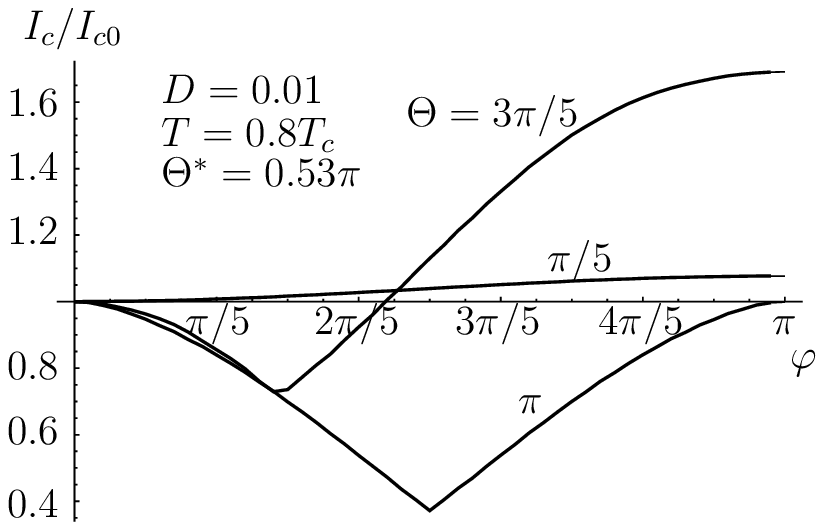,width=\linewidth}
  \end{minipage}\hfill
  \begin{minipage}[b]{.5\linewidth}
   \centering \epsfig{figure=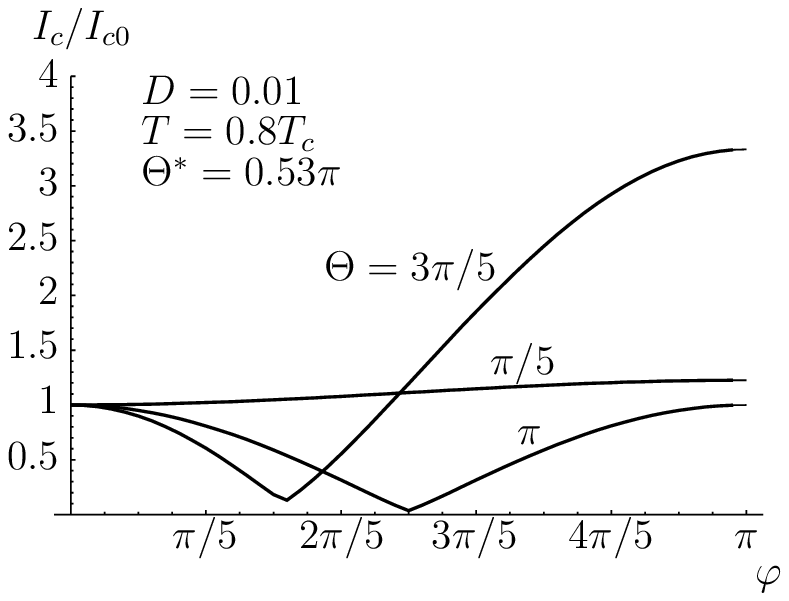,width=\linewidth}
  \end{minipage}\hfill
\caption{ Critical current as a function of the
misorientation angle $\varphi$, normalized to its value
at $\varphi=0$. In the particular case $\varphi=0$
the $0$-$\pi$ transition takes place for $\Theta>\Theta^*(T,D)$.}
\label{cur}
\end{figure}

{\it The Josephson current} is
carried by
the bound states (\ref{cd}),
analogously to the situation in nonmagnetic symmetric
junctions \cite{fur90,fur191,been191,been291}. Hence, in a quantum point
contact with a {\it FIF} constriction the total Josephson current
carried by four Andreev states (\ref{cd}) can be found as
$J(\chi,T)=2e\sum \limits_m \frac{\displaystyle d
\varepsilon_m}{\displaystyle d \chi}n(\varepsilon_m)$ $= -2 e \sum
\limits_{\varepsilon_m>0} \frac{\displaystyle d
\varepsilon_m}{\displaystyle d \chi}{\rm tanh}\frac{\displaystyle
\varepsilon_m}{\displaystyle 2 T}$\enspace . It is not difficult to
calculate the current in this scheme
numerically. The Josephson critical
current as a function of the misorientation angle $\varphi$, normalized to its
value at $\varphi=0$, is shown in Fig. \ref{cur} for various $\Theta$
and for two values of the transparency $D=0.01,\ 0.8$ (the upper and
the lower panels respectively) and the temperature $T=0.1T_c,\ 0.8T_c$
(the right and the left columns).

We define the critical current as a positive quantity, as it is usually
determined experimentally. There are two qualitatively different
regimes for the behavior of the Josephson critical current as a
function of $\varphi$ in all particular cases displayed in Fig.
\ref{cur}. The two regimes are separated by a characteristic value
$\Theta^*(T,D)$, which depends on the temperature and the transparency.
For $\Theta<\Theta^*$ the current is a monotonous function of the
misorientation angle, reaching the maximum for the antiparallel
orientation of the magnetizations. For $\Theta>\Theta^*$ the current
manifests, however, a nonmonotonic dependence on the misorientation
angle with well pronounced minimum at some intermediate value of
$\varphi$ and the maximum at $\varphi=\pi$. In the case $\Theta=\pi$
the currents at $\varphi=0$ and $\varphi=\pi$ are equal to each other.
The parameter $\Theta^*$ admits a simple physical interpretation,
associated with the junction at $\varphi=0$. At zero misorientation
angle the junction acquires a uniformly oriented ferromagnetic interface.
Then the Josephson current is equivalent to that studied
in \cite{fogel00,fogel01,nazarov01,bb02}. It can be shown, that for
$\varphi=0$ and $\Theta=\Theta^*(T,D)$ the $0-\pi$ transition takes
place in the junction just at the given temperature $T$. Hence, for
$\Theta>\Theta^*(T,D)$ the equilibrium state of the junction with
$\varphi=0$ is the $\pi$-state, while for $\Theta<\Theta^*(T,D)$ it is
the $0$-state. We omit an analytical analysis of the total Josephson
current in the case $\varphi=0$, since the results exactly coincide
with those obtained in Refs.~\onlinecite{nazarov01,bb02}.

Furthermore, there is no $0-\pi$ transition in the junction with
antiparallel magnetization, $\varphi=\pi$, in the three-layer
interface. Indeed, it is straightforward to get from Eq.~(\ref{cd}) the
Josephson current in the particular case $\varphi=\pi$:
\begin{eqnarray}
\nonumber
J(\chi,T)=\frac{\displaystyle eD|{\mit\Delta}|\sin\chi}{\displaystyle\sqrt{
\displaystyle D\cos^2\frac{\displaystyle\chi}{\displaystyle2}+R\cos^2\frac{
\displaystyle\Theta}{\displaystyle2}}}\times \qquad \qquad \qquad
\\
\qquad \qquad \qquad
\times\tanh\frac{\displaystyle|{\mit\Delta}|\sqrt{
\displaystyle D\cos^2\frac{\displaystyle\chi}{\displaystyle2}+R\cos^2\frac{
\displaystyle\Theta}{\displaystyle2}}}{\displaystyle2T}
\enspace .
\label{curpi}
\end{eqnarray}
In contrast to the case $\varphi=0$, the current (\ref{curpi}) does not
change its sign at any temperature. The same assertion is valid also
for junctions with dirty superconductors, where Andreev states are
fully destroyed \cite{koshina012,gol021}. We conclude, that the
nonmonotonic dependence of the Josephson current on $\varphi$ arises
due to the $0-\pi$ transition taking place with varying the
misorientation angle at an intermediate value of $\varphi$. This always
occurs under the condition that there is a $\pi$-junction at
$\varphi=0$. If one defined the critical current in the $\pi$-junction
as the negative quantity, then the nonmonotonic behavior would
transform into the monotonic one, accompanied with a change of
sign and discontinuity whenever ${\rm min}\ |J_c|\ne 0$.

The dependence of the Josephson current on the misorientation angle
$\varphi$ becomes especially simple and clear in the tunneling limit.
In tunnel quantum point contacts the Josephson current takes the form
$J(T,\varphi,\chi)=J(T,\varphi)\sin\chi$\ , where
\begin{eqnarray}
&&J(T,\varphi)=J^{(p)}(T)\cos^2\frac{\displaystyle\varphi}{
\displaystyle2}+J^{(a)}(T)\sin^2\frac{\displaystyle\varphi}{\displaystyle
2} \enspace ,
\label{mis} \\
&&J^{(p)}(T)=eD|{\mit\Delta}|\left[
\cos\frac{\Theta}{2}
{\rm tanh}\left(\frac{\displaystyle
|{\mit\Delta}|\cos \frac{\Theta}{2}}{\displaystyle 2
T}\right)-\right. \qquad \quad \nonumber \\ &&
\qquad \qquad \qquad \qquad \quad \left.
-\frac{\displaystyle |{\mit\Delta}|}{\displaystyle 2
T}\frac{\displaystyle \sin^2 \frac{\Theta}{2}}
{\displaystyle \cosh^2
\left(\frac{|{\mit\Delta}|\cos \frac{\Theta}{2}}{2 T}\right)} \right],
\label{tunneltok0}\\
&&J^{(a)}(T)=\frac{\displaystyle
eD|{\mit\Delta}|}{\displaystyle\cos\frac{ \Theta}{2}}{\rm
tanh}\left(\frac{\displaystyle |{\mit\Delta}|\cos
\frac{\Theta}{2}}{\displaystyle 2 T}\right) \enspace .
\label{tunneltokpi}
\end{eqnarray}

The quantities $J^{(p)}(T)$, $J^{(a)}(T)$ are defined as
$J^{(p)}(T)\equiv J(T,\varphi=0)$,\ $J^{(a)}(T)\equiv
J(T,\varphi=\pi)$, so that $|J^{(p)}(T)|$, $|J^{(a)}(T)|$ are critical
currents in tunnel junctions with parallel and antiparallel
orientations of the exchange fields in the three-layer interface. The
dependence (\ref{mis}) on the misorientation angle has been derived in
the tunneling limit earlier in Ref.\,~\onlinecite{volkov01},
disregarding the contribution from Andreev states and, hence, the
$0-\pi$ transition. As one can conclude from (\ref{mis}), the $0-\pi$
transition can take place with varying $\varphi$, if $J^{(p)}_{c}(T)$
and $J^{(a)}_{c}(T)$ have opposite signs. This is exactly the reason
for a nonmonotonic dependence of the critical current on $\varphi$ to
show up. Eq.~(\ref{mis}) is quite general within the tunneling limit
and not applicable to highly transparent junctions. Spin polarizations
of the eigenstates on any side of the impenetrable interface are
aligned parallel or antiparallel to the respective magnetization
direction. Making the projections of the spin polarized states from one
side on the eigenstates on another side (with the spin polarization
rotated by the angle $\varphi$ with respect to the initial one), one
confirms in the tunneling limit that the current is of the form (\ref{mis}).

In conclusion, we have investigated theoretically spectra and spin structures
of interface Andreev states in {\it S-FIF-S} junctions. Both finite
transparency and the misorientation angle between in-plane magnetizations
of ferromagnetic layers were taken into account. We demonstrated that the
Josephson critical current as a function of the misorientation angle always
manifests a nonmonotonic behavior, if at $\varphi=0$ the equilibrium state of
the quantum point contact is the $\pi$-state.

{\it Acknowledgments} This work was supported by the
Russian Foundation for Basic Research under Grant No. 02-02-16643
and by BMBF~13N6918/1.





\end{document}